\documentclass[longauth]{aa} % for the long lists of affiliations 
\usepackage{graphicx}
\usepackage{txfonts}
\usepackage{natbib}
\usepackage{longtable}
\def\ms{\hbox{\,m\,s$^{-1}$}}         %m.s -1
\def\m2s2{\hbox{\,m$^{2}$\,s$^{-2}$}} %m2.s -2
\def\kms{\hbox{\,km\,s$^{-1}$}}       %km.s -1
\def\Msun{\hbox{$M_{\odot}$}}             %Msun
\def\Rsun{\hbox{$R_{\odot}$}}
\def\Mjup{\hbox{$\mathrm{M}_{\rm Jup}$}}
\def\Rjup{\hbox{$\mathrm{R}_{\rm Jup}$}}

\begin{document}

\title{Transiting exoplanets from the CoRoT space mission\thanks{The CoRoT space mission, launched on December 27th 2006, has been developed and is operated by CNES, with the contribution of Austria, Belgium, Brazil, ESA (RSSD and Science Programme), Germany and Spain.}}
\subtitle{XI. CoRoT-10b: a giant planet in a 13.24 day eccentric orbit}

    \titlerunning{CoRoT-10b}
    \authorrunning{A. S. Bonomo et al. 2010}

\author{A.~S.~Bonomo \inst{1} 
\and A.~Santerne\inst{1, 7}
\and R.~Alonso\inst{2}
\and J.-C.~Gazzano\inst{1,3}
\and M.~Havel\inst{3}
%%% Alphabetic list
\and S.~Aigrain\inst{4} 
\and M.~Auvergne\inst{5} 
\and A.~Baglin\inst{5}  
\and M.~Barbieri\inst{21}
\and P.~Barge\inst{1} 
\and W.~Benz\inst{23}
\and P.~Bord\'e\inst{6} 
\and F.~Bouchy\inst{7,8} 
\and H.~Bruntt\inst{5} 
\and J.~Cabrera\inst{9, 19} 
\and A.~C.~Cameron\inst{22}
\and L.~Carone\inst{10} 
\and S.~Carpano\inst{11}
\and Sz.~Csizmadia\inst{9} 
\and M.~Deleuil\inst{1}
\and H.~J.~Deeg\inst{12} 
\and R.~Dvorak\inst{13} 
\and A.~Erikson\inst{9}
\and S.~Ferraz-Mello\inst{14} 
\and M.~Fridlund\inst{11}
\and D.~Gandolfi\inst{11, 15} 
\and M.~Gillon\inst{16} 
\and E.~Guenther\inst{15}
\and T.~Guillot\inst{3} 
\and A.~Hatzes\inst{15} 
\and G.~H\'ebrard\inst{8} 
\and L.~Jorda\inst{1} 
\and H.~Lammer\inst{17}
\and A.~F.~Lanza\inst{20}
\and A.~L\'eger\inst{6} 
\and A.~Llebaria\inst{1} 
\and M.~Mayor\inst{2} 
\and T.~Mazeh\inst{18} 
\and C.~Moutou\inst{1} 
\and M.~Ollivier\inst{6} 
\and M.~P\"atzold\inst{10}
\and F.~Pepe\inst{2} 
\and D.~Queloz\inst{2}
\and H.~Rauer\inst{9, 24} 
\and D.~Rouan\inst{5}
\and B.~Samuel\inst{6}
\and J.~Schneider\inst{19} 
\and B.~Tingley\inst{12} 
\and S.~Udry\inst{2}
\and G.~Wuchterl\inst{15}}

% AGGIUNGERE: Barbieri, Lanza, Cameron

\institute{
Laboratoire d'Astrophysique de Marseille, Universit\'e Aix-Marseille \& CNRS, 38 rue Fr\'ed\'eric Joliot-Curie, F-13388 Marseille Cedex 13, France
\and Observatoire de l'Universit\'e de Gen\`eve, 51 chemin des Maillettes, 1290 Sauverny, Switzerland 
\and Observatoire de la C\^ote dÕ Azur, Laboratoire Cassiop\'ee, BP 4229, 06304 Nice Cedex 4, France
\and Department of Physics, Denys Wilkinson Building Keble Road, Oxford, OX1 3RH
\and LESIA, UMR 8109 CNRS , Observatoire de Paris, UVSQ, Universit\'e Paris-Diderot, 5 place J. Janssen, 92195 Meudon, France
\and Institut d'Astrophysique Spatiale, Universit\'e Paris-Sud 11 \& CNRS (UMR 8617), B\^at. 121, 91405 Orsay, France
\and Observatoire de Haute-Provence, UniversitŽ Aix-Marseille \& CNRS, F-04870 St.~Michel l'Observatoire, France 
\and Institut d'Astrophysique de Paris, UMR7095 CNRS, Universit\'e Pierre \& Marie Curie, 98bis boulevard Arago, 75014 Paris, France
\and Institute of Planetary Research, German Aerospace Center, Rutherfordstrasse 2, 12489 Berlin, Germany
\and Rheinisches Institut f\"ur Umweltforschung an der Universit\"at zu K\"oln, Aachener Strasse 209, 50931, Germany 
\and Research and ScientiÞc Support Department, ESTEC/ESA, PO Box 299, 2200 AG Noordwijk, The Netherlands 
\and Instituto de Astrofõsica de Canarias, E-38205 La Laguna, Tenerife, Spain 
\and University of Vienna, Institute of Astronomy, T\"urkenschanzstr. 17, A-1180 Vienna, Austria
\and IAG, University of Sao Paulo, Brazil 
\and Th\"uringer Landessternwarte, Sternwarte 5, Tautenburg 5, D-07778 Tautenburg, Germany
\and University of Li\`ege, All\'ee du 6 ao\^ut 17, Sart Tilman, Li\`ege 1, Belgium
\and Space Research Institute, Austrian Academy of Science, Schmiedlstr. 6, A-8042 Graz, Austria 
\and School of Physics and Astronomy, Raymond and Beverly Sackler Faculty of Exact Sciences, Tel Aviv University, Tel Aviv, Israel  
\and LUTH, Observatoire de Paris, CNRS, Universit\'e Paris Diderot; 5 place Jules Janssen, 92195 Meudon, France
\and INAF, Osservatorio Astrofisico di Catania, Via S. Sofia, 78, 95123 Catania, Italy
\and Dipartimento di Astronomia, Universit\`a di Padova, 35122 Padova, Italy
\and SUPA, School of Physics and Astronomy, University of St Andrews, Fife KY16 9SS 
\and Universit\"at Bern Physics Inst, Sidlerstrasse 5, CH 3012 Bern, Switzerland 
\and Center for Astronomy and Astrophysics, TU Berlin, Hardenbergstr. 36, 10623 Berlin, Germany }

%\author{A.~S.~Bonomo\inst{1}, A. Santerne\inst{1}, R. Alonso\inst{2}, 
 %    J.-C. Gazzano\inst{1}, M. Havel\inst{3}
%     et al. }

% Barbieri, Lanza, Cameron

%\offprints{A.~S.~Bonomo}

%\institute{ LAM \\
%	\email{aldo.bonomo@oamp.fr} \and
%	Observatoire de Geneve \and Observatoire de la Cote d'Azur} 

%   \date{Received September 15, 1996; accepted March 16, 1997}
	\date{Received ?; accepted ?}

\offprints{A.~S.~Bonomo\\
 \email{aldo.bonomo@oamp.fr}}

\abstract{The space telescope CoRoT searches for transiting extrasolar
planets by continuously monitoring the optical flux of thousands of stars
in several fields of view.}
{We report the discovery of CoRoT-10b, a giant planet on a highly eccentric orbit ($e=0.53 \pm 0.04$)
revolving in 13.24 days around a faint (V=15.22) metal-rich K1V star.}
{We use CoRoT photometry, radial velocity observations taken with 
the \emph{HARPS} spectrograph, and \emph{UVES} spectra of the parent star to
derive the orbital, stellar and planetary parameters.}
{We derive a radius of the planet of $0.97 \pm 0.07~\Rjup$ and a mass 
of $2.75 \pm 0.16~\Mjup$. The bulk density, 
$\rho_{\rm p}=3.70 \pm 0.83 ~\rm{g\;cm^{-3}}$, is $\sim 2.8$ that of Jupiter.
The core of CoRoT-10b could contain up to 
240 M$_\oplus$ of heavy elements. \\
Moving along its eccentric orbit,
the planet experiences a 10.6-fold variation in insolation. 
Owing to the long circularisation time,  $\tau_{\rm circ} > 7$~Gyr, a resonant perturber 
 is not required to excite and maintain the high eccentricity of CoRoT-10b.}
{}

	
% 5 {} token are mandatory
 
 \maketitle
%
%________________________________________________________________

%	\abstract{CoRoT is the pioneer space mission dedicated to the detection of extrasolar
%planets via the transit method. It continueously monitors the optical flux of thousands of stars
%in each of its fields of view.}


\section{Introduction}
CoRoT is the pioneer space mission dedicated to the 
detection of extrasolar 
planets via the transit method (\citealt{Baglin03}; see \citealt{Auvergneetal09} 
for a detailed description of the instrument and its performance). 
To date it has led to the discovery of 
ten extrasolar planets: CoRoT-7b, the first super-Earth with 
measured radius and mass \citep{Legeretal09, Quelozetal09}; three inflated hot Jupiters, 
CoRoT-1b \citep{Bargeetal08}, CoRoT-2b \citep{Alonsoetal08a} and CoRoT-5b \citep{Raueretal09}; 
CoRoT-3b, ``the 
first secure inhabitant of the brown dwarf desert'' \citep{Deleuiletal08}; 
two Jupiter-like planets with an orbital period of 
approximately 9 days, CoRoT-4b \citep{Aigrainetal08, Moutouetal08} and 
CoRoT-6b \citep{Fridlundetal10}; the long-period temperate giant
 planet CoRoT-9b \citep{Deegetal10}, and the hot Super-Neptune
 CoRoT-8b (\citealt{Bordeetal10}).

Here we report the discovery of the giant planet CoRoT-10b that orbits
its parent star in 13.24 days, moving along a highly
eccentric orbit with $e=0.53 \pm 0.04$. It is therefore one of the few known
transiting planets with $e \gtrsim 0.5$, such as
HD~147056b (alias HAT-P-2b, $e=0.52$ and orbital period $P=5.63$ days;
\citealt{Bakosetal07, Paletal10}),
HD~17156b ($e=0.67$ and $P=21.22$ days; \citealt{Barbierietal07, Barbierietal09}) and 
HD~80606b ($e=0.93$ and $P=111.44$ days ; \citealt{Moutouetal09, Winnetal09, Hebrardetal10}).
Eccentric transiting planets include also those with 
a smaller eccentricity, notably
the recently discovered WASP-8b ($e=0.31$ and 
$P=8.16$ days; \citealt{Quelozetal10}) and HAT-P-15b 
($e=0.19$ and 
$P=10.9$ days; \citealt{Kovacsetal10}), 
the massive planet X0-3b ($e=0.26$ and $P=3.19$ days; 
\citealt{JohnsKrulletal08, Winnetal08}) and the two Neptunes
GJ~436b ($e=0.15$ and $P=2.64$ days; \citealt{Gillonetal07a, Gillonetal07b,
Alonsoetal08b, Beanetal08}) and 
HAT-P-11b ($e=0.20$ and $P=4.89$ days; \citealt{Bakosetal10}).

Transiting planets in eccentric orbits are very intriguing and
interesting objects as they allow us to study 
ongoing tidal dissipation and its impact 
on the planet radius \citep{Ibgui10},
atmospheric circulation in the case of a strong
variation in insolation \citep{Langton07}, and the
dynamical orbital evolution including the 
gravitational interaction between planets
in a multiple system (planet-planet scattering; e.g., \citealt{Marzari02}) 
or the secular influence of a possible distant stellar 
companion (Kozai mechanism; \citealt{Kozai62}).

\section{CoRoT observations}
\label{corot_obs}
%The parent star of CoRoT-10b, i.e. the CoRoT target LRc01\_E2\_1802, is 
%a $V=15.22$ star and has been 
%observed in the LRc01 stellar field pointing towards the constellation of 
%Aquila \citep{Cabreraetal09}. 

The parent star of CoRoT-10b, i.e. the CoRoT target LRc01\_E2\_1802, is 
a $V=15.22$ star and has been 
observed in the stellar field pointing towards the constellation of 
Aquila during the first CoRoT long run LRc01 \citep{Cabreraetal09}. 
Its magnitudes in several 
photometric bands and its coordinates are reported in Table~\ref{startable}.
CoRoT observations of this target lasted for 142.07 days, from the $16^{\rm{th}}$ of 
May up to the $15^{\rm{th}}$ of October 2007, and
provided us with monochromatic (white channel) data \citep{Auvergneetal09}.

\begin{figure}[t]
\centering
\includegraphics[width=6.5cm, angle=90]{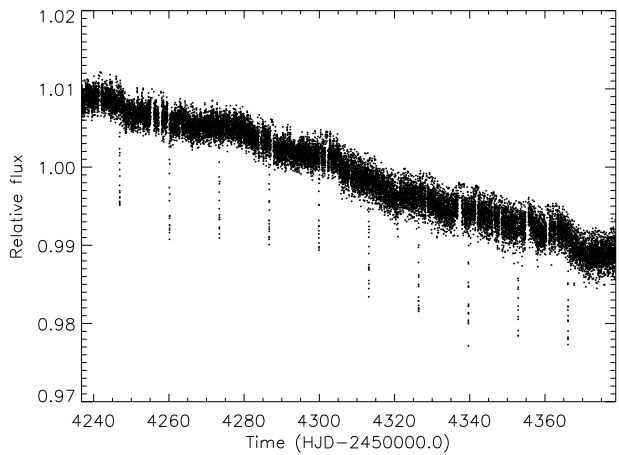}
\caption{The quiescent light curve of CoRoT-10 binned at 512~s
showing ten transits of the giant planet CoRoT-10b.
Jumps due to hot pixels were removed by means
of an iterative 3-sigma clipping.}
\label{lcfig}
\end{figure}

\begin{table}
\caption{CoRoT-10 IDs, coordinates and magnitudes.}            
%\begin{minipage}{3.5 cm} 
\centering        
\begin{minipage}[!]{7.0cm}  
\renewcommand{\footnoterule}{}     
\begin{tabular}{lcc}       
\hline\hline                 
%BJD & RV & $\pm$$1\,\sigma$ & exp. time & S/N p. pix. \\
%-2\,400\,000 & (km\,s$^{-1}$) & (km\,s$^{-1}$) & (sec) &  (at 550 nm)  \\
CoRoT window ID & LRc01\_E2\_1802 \\
CoRoT ID & 100725706 \\
USNO-A2 ID  & 0900-14919216 \\
2MASS ID   & 19241528+0044461 \\
GSC2.3 ID & NIMR021985 \\
\\
\multicolumn{2}{l}{Coordinates} \\
\hline            
RA (J2000)  & 19:24:15.29 \\
Dec (J2000) & 00:44:46.11 \\
\\
\multicolumn{3}{l}{Magnitudes} \\
\hline
\centering
Filter & Mag & Error \\
\hline
B$^a$  & 16.68 & 0.14 \\
V$^a$  & 15.22 & 0.05 \\
r'$^a$ & 14.73 & 0.03 \\
i'$^a$ & 13.74 & 0.03 \\
J$^b$  & 12.53 & 0.02 \\
H$^b$  & 11.93 & 0.03 \\
K$^b$  & 11.78 & 0.02 \\
%\\                                    
%\multicolumn{3}{l}{Proper motion} \\
%\hline
%$\mu_{\alpha}$ & 8.0   & ''/yr
%$\mu_{\alpha}$ & -11.8 & ''/yr
\hline\hline
\vspace{-0.5cm}
\footnotetext[1]{Provided by Exo-Dat \citep{Deleuiletal09};}
\footnotetext[2]{from 2MASS catalogue.}
\end{tabular}
\end{minipage}
\label{startable}      
\end{table}

Transits by CoRoT-10b were first discovered in ``alarm mode'' \citep{Suraceetal08}, 
i.e. while CoRoT observations were still ongoing, 
which permitted us to change the 
temporal sampling from 512~s to 32~s after HJD 2454305.11. 
In total, 210248 photometric measurements were obtained, 
198752 in the 32~s oversampling mode \footnote{data available at
http://idoc-corot.ias.u-psud.fr/}. 
Fig.~\ref{lcfig} shows the CoRoT-10 light curve with the nominal 
sampling of 512~s, filtered from a) outliers that are 
produced by proton impacts during the crossing of
the South Atlantic Anomaly of the Earth's magnetic 
field by the satellite;
% and represent $7.2 \%$
%of the data points
and b) several jumps, with a typical duration
shorter than 1~day, due to hot pixels . 
Corrections for the CCD 
zero offset, sky background, Earth's scattered 
light and jitter variations were carried out 
by the latest version 2.1 of the CoRoT reduction pipeline.
Unlike CoRoT-2 \citep{Lanzaetal09}, CoRoT-6 \citep{Fridlundetal10} and 
CoRoT-7  (\citealt{Lanzaetal10}), 
the light curve of CoRoT-10 is relatively quiescent and 
does not show flux variations due to the presence of starspots 
and photospheric faculae greater than a few mmag. 
It shows a long term decrease of $\sim 2.5 \%$ attributable to 
CCD ageing (Fig.~\ref{lcfig}).
The rms of the nominal and
oversampled photometric points is $0.0013$ and $ 0.0046$ in relative 
flux units, respectively.

\begin{figure}[b!]
\centering
\includegraphics[width=6.5cm, angle=90]{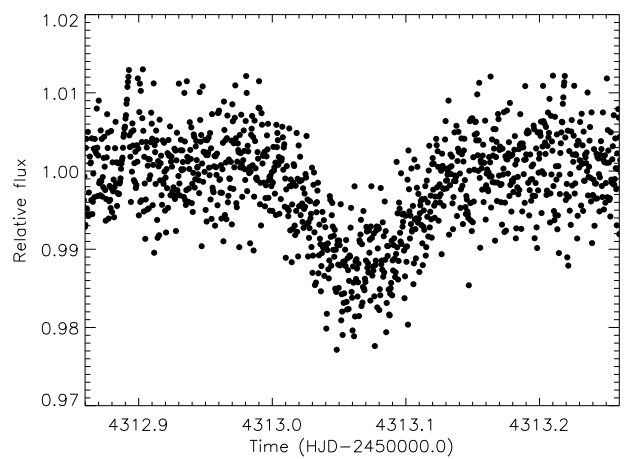}
\caption{One of the five 32~s oversampled transits of CoRoT-10b.}
\label{1transitfig}
\end{figure}

A total of 10 transits with a depth of $\sim 1.3 \%$
are visible in the light curve
(Fig.~\ref{lcfig}). 
A zoom of one of the five 32~s oversampled transits is shown in 
Fig.~\ref{1transitfig}.
The transit ephemeris, reported in Table~\ref{starplanet_param_table}, 
was derived from a linear fit to the
measured transit mid points determined by a trapezoidal fitting of
each transit. It gives an orbital period
of $13.2406 \pm 0.0002$ days. 

\section{Ground-based follow-up observations}

\subsection{Radial velocity observations}
\label{RV_obs}
We performed radial velocity (RV) observations of the star 
CoRoT-10 with the \emph{HARPS} 
spectrograph (\citealt{pepe02b}, \citealt{mayor03}) at
the 3.6-m ESO telescope (La Silla, Chile). 
\emph{HARPS} was used with the observing mode obj\_AB, without 
acquisition of a simultaneous Thorium lamp spectrum
in order to monitor the 
Moon background light on the second fibre. 
The intrinsic stability of this spectrograph does not require the use of 
lamp calibration spectra, the instrumental drift during one night being in our 
case always smaller than the stellar RV photon noise uncertainties.  
\emph{HARPS} data were reduced with the on-line standard pipeline and radial velocities 
were obtained by a weighted cross-correlation with a 
numerical spectral mask for a K5V star
(\citealt{baranne96};  \citealt{pepe02a}). 

\begin{figure}[t]
%\centering
\includegraphics[width=8.5cm]{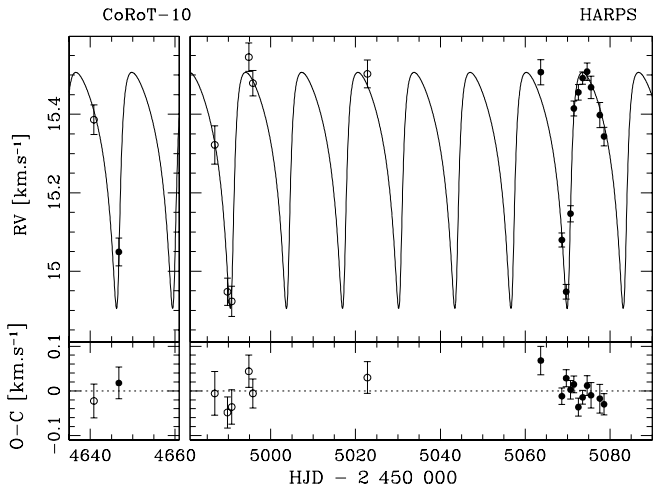}
\caption{\emph{Top panel}: radial velocity measurements obtained
 and the Keplerian best-fit solution (solid line).
\emph{Bottom panel}: residuals from the best-fit.
The open circles indicate the measurements
affected by the moonlight after our correction.}
\label{rvvstime}
\end{figure}

The first two measurements of CoRoT-10 were made on June 2008 
\footnote{\emph{HARPS} program 081.C-0388}. 
Seventeen additional measurements were obtained from June to September 
2009\footnote{\emph{HARPS} program 083.C-0186}. 
Seven of our nineteen measurements were strongly contaminated by the moonlight: 
the radial velocity of the Moon was close to that of CoRoT-10 
and affected both the RV measurements and the bisector lines. 
We developed a software correction using the Moon spectrum 
simultaneously acquired on fibre B: it consists in subtracting the cross-correlation 
function (CCF) of fibre B, containing the Sun spectrum (reflected
by the Moon), from the CCF 
of the fibre A, containing the stellar spectrum. 
The correction was applied when the two CCF peaks 
were close in radial velocity. For CoRoT-10, corrections in the range 
between 50 and 550 {\ms} were applied for seven measurements. 
To be conservative, we added quadratically 30 {\ms} of systematic errors 
to these seven  corrected measurements. 

\begin{figure}[!h]
\centering
\includegraphics[width=8.5cm]{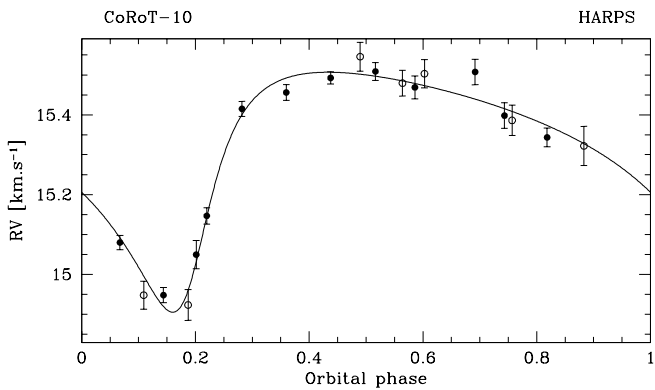}
\caption{Phase-folded radial velocity curve of CoRoT-10 and the 
Keplerian best-fit solution (solid line). The open circles 
indicate the measurements
affected by the moonlight after our correction.}
\label{rvphase}
\end{figure}

The radial velocities are listed in Table~\ref{rvtable} and displayed 
in  Fig.~\ref{rvvstime} and ~\ref{rvphase}. 
The phase-folded radial velocity curve shows a variation 
in phase with the CoRoT transit period. It is compatible with the reflex 
motion of the parent star due to an eccentric planetary 
companion. We fitted the data with a 
Keplerian orbit using the 
CoRoT ephemeris $P=13.2406$ days and $T_{tr}=2454273.3436$ HJD
(see Table~\ref{starplanet_param_table}). 
The derived eccentricity and argument of
periastron are $e=0.53 \pm 0.04$ and 
$\omega=218.9 \pm 6.4$~deg.
The other orbital parameters 
are reported in Table~\ref{starplanet_param_table}. 
The standard deviation of the residuals to the fit
$\sigma(O-C)=29 $~m\,s$^{-1}$ is comparable to the mean 
RV uncertainty.

\begin{figure}[h!]
%\centering
\includegraphics[width=8.5cm]{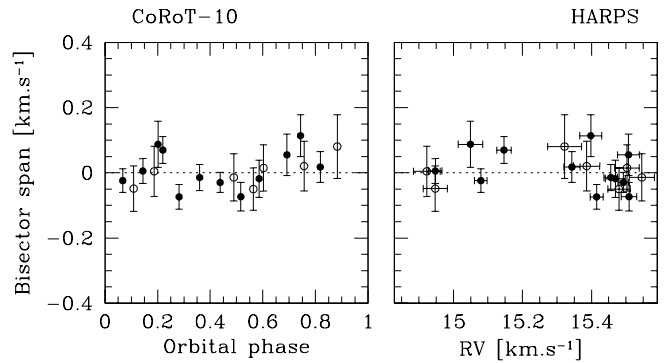}
\caption{Bisector variations (span of the bisector slope) as a function of orbital phase
 (\emph{left pannel}) and radial velocity (\emph{right pannel}). Bisector error bars are 
 estimated as twice the radial velocity uncertainties. No bisector effect is visible for the 
 moonlight-corrected measurements (open circles) indicating the good quality 
 of our correction.}
\label{bisector}
\end{figure}

To examine the possibility that the RV variation is due to a 
blended binary scenario -- a single star with an unresolved 
and diluted eclipsing binary --, 
we followed the procedure described in \citet{bouchy08} 
based on checking 
both the spectral line asymmetries and the dependences of the RV variations on
different cross-correlation masks. These two checks excluded 
that the RV variation is caused by a blended binary and allowed us to establish the planetary 
nature of CoRoT-10b. The bisector variations are shown in Fig.~\ref{bisector}.

\begin{table}
\caption{Radial velocity measurements of CoRoT-10 obtained by \emph{HARPS}. 
HJD is the Heliocentric Julian Date. The seven exposures which were affected by the moonlight 
are labelled with an asterisk  and 30 {\ms} was quadratically added to their errors. }            
\centering                          
\begin{minipage}{9cm}
\setlength{\tabcolsep}{1.2mm}
%\begin{tabular}{llllll}   
\begin{tabular}{lccccc}   
\hline\hline                 
HJD & RV & $\pm$$1\,\sigma$ & BIS & exp. time & S/N/pix. \\
-2\,400\,000 & [km\,s$^{-1}$] & [km\,s$^{-1}$] & [km\,s$^{-1}$] & [s] &  (at 550nm)  \\
\hline 
54640.85815$^ *$  & 15.386  & 0.038 & 0.021 &  3600 &  7.4  \\
54646.74608          & 15.049  & 0.035          & 0.087 &  1800 &  5.1  \\
54986.78569$^ *$  & 15.322  & 0.049 & 0.081 &  3600 &  4.5  \\ 
54989.77932$^ *$  & 14.947  & 0.035 & -0.048 &  3600 &  9.2  \\
54990.81302$^ *$  & 14.923  & 0.038 & 0.004 &  3600 &  7.1  \\
54994.81853$^ *$  & 15.545  & 0.036 & -0.014 &  3600 &  8.4  \\
54995.80369$^ *$  & 15.479  & 0.032 & -0.050 &  3600 &  10.8 \\
55022.79584$^ *$  & 15.503  & 0.035 & 0.015 &  3600 &  7.0  \\
55063.69761          & 15.507  & 0.032 	     & 0.055 &  3300 &  5.4  \\
55068.67093          & 15.079  & 0.018          & -0.024 &  3600 &  7.7  \\
55069.68356  	   & 14.947  & 0.019          & 0.006 &  3300 &  7.6  \\
55070.69173  	   & 15.146  & 0.021          & 0.070 &  3000 &  7.4  \\
55071.51030          & 15.414  & 0.019          & -0.074 &  3000 &  7.8  \\
55072.54251          & 15.456  & 0.020          & -0.014 &  3000 &  7.5  \\
55073.57439          & 15.492  & 0.015          & -0.030 &  3000 &  9.1  \\
55074.61800          & 15.508  & 0.022          & -0.073 &  3600 &  7.7  \\
55075.53585          & 15.468  & 0.029          & -0.018 &  3000 &  6.4  \\
55077.62048          & 15.398  & 0.032          & 0.114 &  3000 &  5.7  \\
55078.61399          & 15.343  & 0.024          & 0.018 &  3600 &  7.6  \\
\hline                                   
\end{tabular}
\label{rvtable}      
\end{minipage} 
\end{table}

\subsection{Photometric observations}
\begin{figure}[h!]
\vspace{-0.5cm}
\centering
\includegraphics[width=6.5cm, angle=-90]{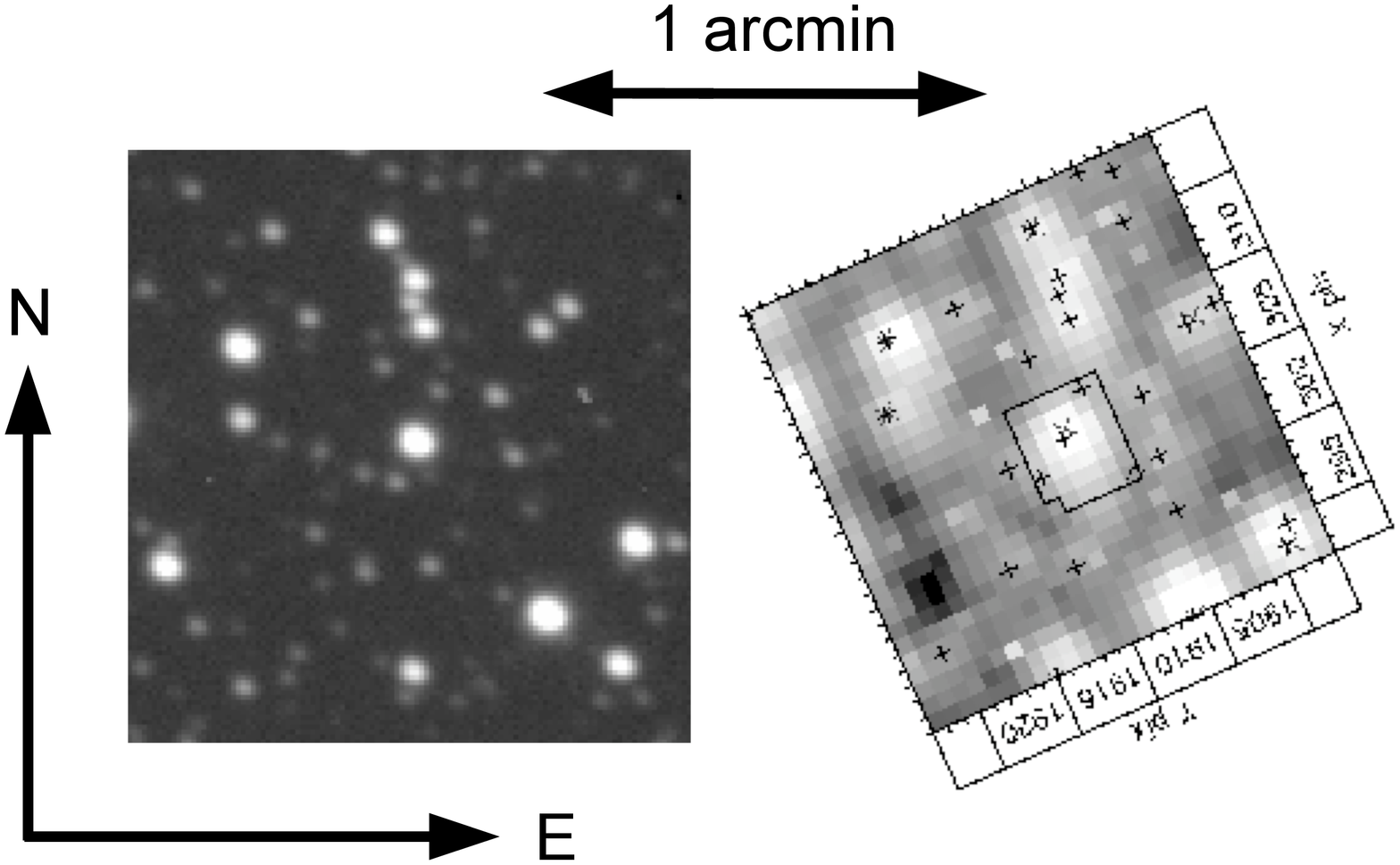}
\caption{The sky area around CoRoT-10 (the brightest star near the centre).
\emph{ Left}: R-filter image with a resolution of $\sim 2.5''$ 
taken with the OHP 1.20~m telescope. \emph{Right}: image taken
by CoRoT, at the same scale and orientation. The jagged outline
in its centre is the photometric aperture mask; indicated are also
CoRoT's x and y image coordinates and position of nearby stars
from the Exo-Dat database \citep{Deleuiletal09}.}
\label{contamfig}
\end{figure}

Photometric observations during and outside the transit 
were carried out at the 1.20 m telescope of the 
Observatoire de Haute Provence 
during the nights of the $\rm{18^{th}}$ of June and the 
$\rm{15^{th}}$ of
July 2009, respectively. Such observations are
complementary to the radial velocity measurements 
and are required
to definitively exclude the possibility that the 
transits detected in the CoRoT light curve could
be produced by a background eclipsing binary 
which contaminates the 
CoRoT aperture mask of the star \citep{Deegetal09}. 
The latter covers  $\sim 16'' \times 14''$ on the sky
and contains two main background contaminants: 
one is 3.78 mag fainter
in V and is located at $\sim 6.5''$ from CoRoT-10 toward South-West;
the other one is 4.22 mag fainter in V and is 
$\sim 7.5''$ toward North-East (see Fig.~\ref{contamfig}).
Differential photometry shows 
an ``on-target" transit with the same depth as observed 
in the CoRoT light curve. None of the two contaminants
exhibits significant flux variations that can mimic
 the transits observed in the CoRoT-10 light curve.

\section{Transit fitting}
\begin{figure*}[t]
\centering
\includegraphics[width=9.5cm, angle=90]{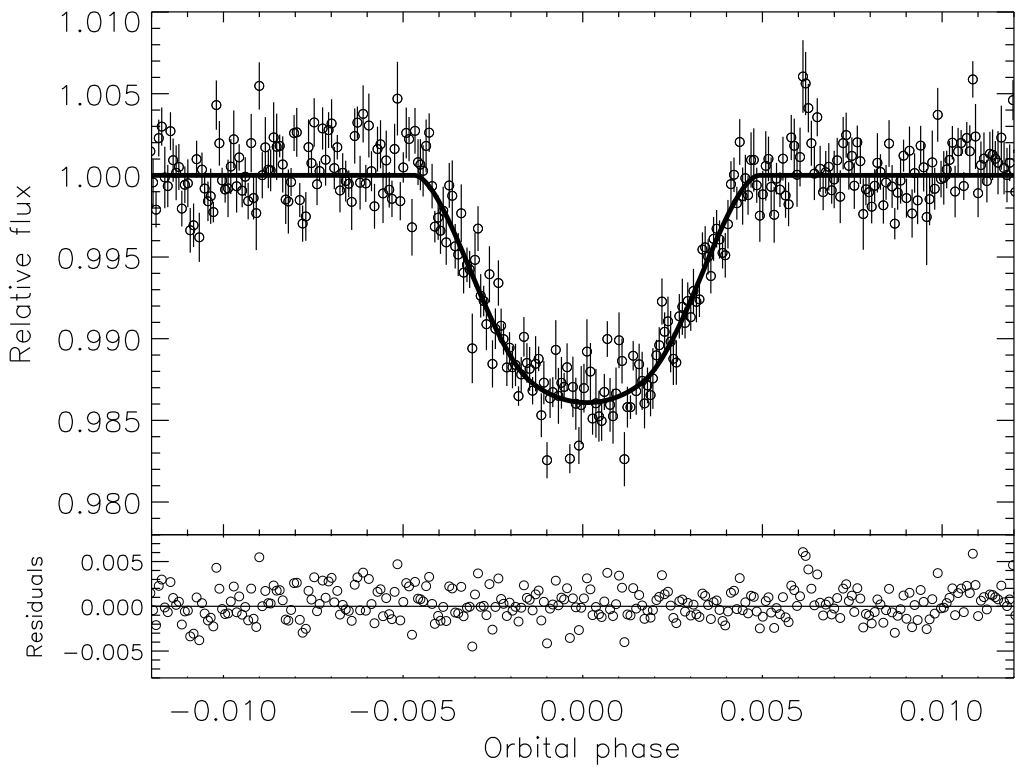}
\hspace{1.0cm}
\vspace{1.0cm}
\caption{\emph{Top panel}: Phase-folded light curve of 8 transits of CoRoT-10b. 
The bin size corresponds to 1.5 min and the 1-sigma error bars
on each bin are estimated as the standard error 
of the data points inside the bin.
The solid line shows our best-fit transit model.
\emph{Bottom panel}: the residuals from the best-fit model.}
\label{tr_bestfit_fig}
\end{figure*}

In order to perform the transit fitting, first of all
we filtered the raw light curve 
from outliers due to impacts of cosmic rays.
Based on pre-launch observations stored in the Exo-Dat catalogue
 \citep{Deleuiletal09}, we estimated the flux contamination from the two faint background
stars which fall inside the CoRoT-10 photometric mask 
to be $5.5 \pm 0.3 \%$ (see Fig.~\ref{contamfig}). We subtracted such a value from
the median flux of the light curve (75559 $e^{-}$/32~s), which makes the transits
slightly deeper by $\sim 7 \cdot 10^{-4}$ in relative flux.
We then fitted a parabola to the 5~h intervals of the light curve before 
the ingress and after the egress of 
each transit in order to correct for any local variations.
We disregarded two of the ten CoRoT-10b transits,
precisely the second transit and the ninth, as their shape was deformed 
by hot pixels. %(see Fig.~\ref{2transitfig}) 
Finally, we folded the 
light curve using the 
ephemeris reported in Table~\ref{starplanet_param_table} and binning 
the data points in bins of
$8 \cdot 10^{-5}$ in phase, corresponding to $\sim 1.5$ min 
(Fig.~\ref{tr_bestfit_fig}). The error on 
each bin was computed as the standard error of the data points
inside the bin. 

%\begin{figure}[h]
%\centering
%\includegraphics[width=8.0cm]{plot_2transit_corot10.ps}
%\hspace{1.0cm}
%\caption{The two transits of CoRoT-10b that have been discarded for the 
%construction of the transit light curve shown in Fig.~\ref{tr_bestfit_fig} as 
%their shape is deformed by hot pixels. One transit was observed 
%with the 512~s nominal sampling (\emph{upper panel}) while
%the other one in the 32~s oversampling mode (\emph{bottom panel}). 
%}
%\label{2transitfig}
%\end{figure}

Transit best-fit was performed following the formalism 
of \citet{Gimenez06, Gimenez09} and
fixing the eccentricity and the argument of the
periastron to the values derived from the Keplerian fit 
of the radial velocity measurements (see Sect.~\ref{RV_obs}). 
The four free parameters of the transit model
are: the transit centre; the phase of the end of transit egress  $\theta_{2}$
in the reference system for eccentric orbits defined by 
\citet{GimenezPelayo83} and 
shown in their Fig.~1 (see also \citealt{Gimenez09});
%the true anomaly measured from conjunction at the transit
%ingress (or egress) $\zeta_{1}$ (see Gim09 for details);
the ratio of the planet to stellar radii $k=R_{\rm p}/R_{*}$, and the inclination $i$ between
the orbital plane and the plane of the sky.
The two non-linear limb-darkening coefficients $u_{+}=u_{a}+u_{b}$
and $u_{-}=u_{a}-u_{b}$ \footnote{$u_{a}$ and $u_{b}$ are the coefficients
of the limb-darkening quadratic law:
$I(\mu)/I(1)=1-u_{a}(1-\mu)-u_{b}(1-\mu)^2$, where $I(1)$ is the 
specific intensity at the centre of the disk and $\mu=\cos{\gamma}$, 
$\gamma$ being the angle between the surface normal and the line of sight} 
were fixed for two reasons: 
first, the relatively low signal-to-noise ratio of the transit
light curve, which does not allow us to constrain either $u_{+}$ 
or $u_{-}$ within reasonable error bars; 
secondly, the degeneracy between the inclination and 
the two limb darkening coefficients in the case of a high impact
parameter $b$, as in our case $b=0.85 \pm 0.03$ 
(see Table~\ref{starplanet_param_table}).
The adopted limb-darkening coefficients $u_{a}$ and $u_{b}$ for 
the CoRoT bandpass were determined following the procedure in
 \citet{Sing10}. However, while the latter takes into account only 
the stellar emergent intensity for the values of
$\mu \ge 0.05$, we considered
all the 17 values available in the ATLAS model grids 
\footnote{http://kurucz.harvard.edu} down to $\mu = 0.01$. 
Our choice is motivated by the fact that the transit of CoRoT-10b is
grazing, which implies that the variation of the specific 
intensity close to the limb of the
stellar disc must be considered properly when modelling
the transit shape. The derived limb-darkening quadratic coefficients 
are $u_{a}=0.51$ and $u_{b}=0.21$, which 
give $u_{+}=0.72$ and $u_{-}=0.3$.

% and the linear limb-darkening 
% coefficient $u$ \footnote{$I(\mu)/I(1)=1-u(1-\mu)$, where $I(1)$ is the 
% specific intensity at the center of the disk and $\mu=\cos{\gamma}$, 
%$\gamma$ beeing the angle between the surface normal and the line of sight;}. 
%We decided to use a simple limb-darkening linear law, 
%after trying to fit the two non-linear
%limb-darkening coefficients $u_{+}=u_{a}+u_{b}$
%and $u_{-}=u_{a}-u_{b}$ \footnote{$u_{a}$ and $u_{b}$ are the coefficients
%of the limb-darkening quadratic law:
%$I(\mu)/I(1)=1-u_{a}(1-\mu)-u_{b}(1-\mu)^2$.}, 
%for two reasons: 
%first, the relatively low signal-to-noise ratio of the transit
%light curve, which does not allow us to constrain either $u_{+}$ 
%or $u_{-}$ within reasonable error bars; 
%secondly, the degeneracy between the inclination and 
%the two limb darkening coefficients in the case of a high impact
%parameter (ref.?). Indeed, in our case
%this last is $b=0.83 \pm 0.02$ (see Table~\ref{starplanet_param_table}). 

The best-fit transit parameters were found
by using the algorithm AMOEBA \citep{Pressetal92} and changing the initial
values of the parameters with a Monte-Carlo method 
to find the global minimum of the $\chi^{2}$.
Our best-fit of the phase-folded 
and binned transit light curve 
is shown in Fig.~\ref{tr_bestfit_fig}.
Fitted and derived transit parameters are listed in
Table~\ref{starplanet_param_table} together with their 1-sigma errors
estimated using the bootstrap procedure 
described in \citet{Alonsoetal08a} which
takes also the correlated noise into account 
(cf. \citealt{Alonsoetal08a}, Sect.~3).
Uncertainties on the eccentricity, the 
argument of the periastron and the contamination
were also considered for the estimation of the errors of
the derived transit parameters $b$, 
$a/R_{*}$, $a/R_{\rm p}$, $M^{1/3}_{*}/R_{*}$ and $\rho_{*}$,  
where $b$ is the impact parameter, $a$ the semi-major axis of the planetary orbit,
$R_{*}$ and $M_{*}$ the stellar radius and mass, 
$R_{\rm p}$ the planet radius, and $\rho_{*}$ the stellar
density (see Table~\ref{starplanet_param_table}).
The fitted value of the transit centre
is consistent with zero within 1-sigma, which confirms the good quality of 
the transit ephemeris (Sect.~\ref{corot_obs}).

%The fitted limb darkening coefficient $u$ is within 3-sigma
%from the theoretical value expected for the CoRoT bandpass \citep{Sing10}.

\section{Stellar and planetary parameters}
\label{stellar_planet_param}

%\subsection{Spectroscopic parameters of the parent star CoRoT-10}
The spectral analysis of the parent star was performed with a high-resolution \emph{UVES} spectrum
acquired on the 30$^{th}$ of July  2009\footnote{\emph{UVES} program ID: 083.C-0690(A)}. 
We used the Dic1 mode (390+580) and a slit width of 0.8'', 
achieving a resolving power of~$\sim$55\,000. The total exposure 
time was 4~h leading to a signal-to-noise ratio per pixel S/N$\sim$120 at 5500~\AA. 

To derive the stellar atmospheric parameters, we first determined 
the $V \sin{i_{*}} = 2\pm0.5$\,km\,s$^{-1}$. To that purpose, we selected a few \emph{HARPS} 
spectra that were not contaminated by the Moon reflected light. This series 
of spectra was set at rest and co-added.
To carry out the detailed spectral analysis, we made use of the \emph{VWA} 
\citep{Bruntt2009} software package and obtained:
T$_{\rm eff}=5075\pm75$~K, log{\it~g}$=4.65\pm0.10$~cm~s$^{-2}$ and [Fe/H]$=0.26\pm0.07$~dex. 
The surface gravity value was checked with usual indicators: Na \textsc{I}~D lines around 
5890~\AA, Mg~\textsc{I}~b lines, and Ca \textsc{I} lines at 6122, 
6162 and 6439~\AA.
The abundances of several chemical elements are listed in Table~\ref{tab:abund}. 
The elements for which we could only measure a few lines are not reported.

The absence of noticeable emission in the core of the 
Ca~II H \& K lines supports the low magnetic activity 
of CoRoT-10 indicated by its 
quiescent light curve (see Fig.~\ref{lcfig}).

\begin{table}
 \centering
 \caption{Abundances of some chemical elements for the fitted lines in the
  \emph{UVES} spectrum. The abundances refer to the solar value 
  and the last column reports the number of lines used.
 \label{tab:abund}}
% \setlength{\tabcolsep}{3pt} % narrow table: default is tabcolsep = 6pt
% \begin{footnotesize}
\begin{tabular}{lcr}
\hline
\noalign{\smallskip}
       Element & Abundance & No. lines \\
\noalign{\smallskip}
\hline
\noalign{\smallskip}
  {Ca \sc   i} &    $0.21  \pm 0.12$  &   5 \\ 
  {Ti \sc   i} &    $0.38  \pm 0.11$  &  15 \\ 
  {V  \sc   i} &    $0.63  \pm 0.12$  &  10 \\ 
  {Cr \sc   i} &    $0.34  \pm 0.13$  &   7 \\ 
  {Fe \sc   i} &    $0.26  \pm 0.10$  &  77 \\ 
  {Fe \sc  ii} &    $0.26  \pm 0.11$  &   5 \\ 
  {Co \sc   i} &    $0.41  \pm 0.10$  &   7 \\
  {Ni \sc   i} &    $0.35  \pm 0.10$  &  23 \\ 
  {Si \sc   i} &    $0.37  \pm 0.10$  &   8 \\ 

\noalign{\smallskip}
\hline
\end{tabular}
%\end{footnotesize}
\end{table}

%CoRoT-10 is a quiet star. Its low magnetic activity level is 
%supported by both the absence of noticeable emission in the core of the 
%CaII H \& K lines and the low amplitude flux variations of its 
%light curve (see Fig.~\ref{lcfig}).  Both the slow rotation and the 
%absence of the Lithium lines 
%at 6708~\AA~and 6104~\AA~ indicate that CoRoT-10 is a relatively old star. 

\begin{table*}
%\vspace{-0.4cm}
\centering
\caption{Planet and star parameters.}            
%\vspace{1cm}
\begin{minipage}[t]{13.0cm} 
\setlength{\tabcolsep}{10.0mm}
\renewcommand{\footnoterule}{}                          
\begin{tabular}{l l}        
\hline\hline                 
%\\
\multicolumn{2}{l}{\emph{Ephemeris}} \\
\hline
Planet orbital period $P$ [days] & 13.2406 $\pm$ 0.0002 \\
Planetary transit epoch $T_{ \rm tr}$ [HJD-2400000] & 54273.3436 $\pm$ 0.0012 \\
Planetary transit duration $d_{\rm tr}$ [h] & 2.98 $\pm$ 0.06 \\
Planetary occultation epoch $T_{\rm occ}$ $^a$ [HJD-2400000] & 54276.49 $\pm$ 0.41 \\
Planetary occultation duration $d_{\rm occ}$ [h] & 2.08 $\pm$ 0.18 \\
Epoch of periastron $T_{0}$ [HJD-2400000] & 54990.85 $\pm$ 0.08 \\
& \\
%\\
\multicolumn{2}{l}{\emph{Derived parameters from radial velocity observations}} \\
\hline    
Orbital eccentricity $e$  &  0.53 $\pm$ 0.04 \\
Argument of periastron $\omega$ [deg] & 218.9 $\pm$ 6.4 \\ 
Radial velocity semi-amplitude $K$ [\ms] & 301 $\pm$ 10 \\
Systemic velocity  $V_{\rm r}$ [\kms] & 15.330 $\pm$ 0.007 \\
O-C residuals [\ms] & 29 \\
& \\
%\\
\multicolumn{2}{l}{\emph{Fitted and fixed transit parameters}} \\
\hline
$\theta_{2}$~$^b$ & 0.00483 $\pm$ 0.00009 \\
 %\(\mbox{ \boldmath $\theta_{2}$} \) ~$^b$ & 0.00483 $\pm$ 0.00009 \\
Radius ratio $k=R_{\rm p}/R_{*}$ & 0.1269 $\pm$ 0.0038 \\
Inclination $i$ [deg] & 88.55 $\pm$ 0.2 \\
%Linear limb darkening coefficient $u$ & 0.84 $\pm$ 0.06 \\
%& \\
%\\
%\multicolumn{2}{l}{\emph{Fixed transit parameters}} \\
%\hline
$u_{+}$ (fixed) & +0.72 \\
$u_{-}$ (fixed) &  +0.30 \\
& \\
%\\
\multicolumn{2}{l}{\emph{Derived transit parameters}} \\
\hline
$a/R_{*}$~$^c$ & 31.33 $\pm$ 2.15 \\
$a/R_{\rm p}$ & 247 $\pm$ 21 \\
$(M_{*}/\Msun)^{1/3} (R_{*}/\Rsun)^{-1}$ & 1.33 $\pm$ 0.09 \\
Stellar density $\rho_{*}$ [$g\;cm^{-3}$] & 3.32 $\pm$ 0.70\\
Impact parameter $b$~$^d$ & 0.85 $\pm$ 0.03\\
& \\
%\\
\multicolumn{2}{l}{\emph{Spectroscopic parameters of the star}} \\
\hline
Effective temperature $T_{\rm{eff}}$[K] & 5075 $\pm$ 75 \\
Surface gravity log\,$g$ [cgs]&  4.65 $\pm$  0.10 \\
Metallicity $[\rm{Fe/H}]$ [dex]& +0.26 $\pm$ 0.07 \\
Stellar rotational velocity $V \sin{i_{*}}$ [\kms] & 2.0 $\pm$ 0.5 \\
Spectral type & K1V \\
& \\
%\\
\multicolumn{2}{l}{\emph{Stellar and planetary physical parameters}} \\
\hline
Star mass [\Msun] ~$^e$ &  0.89 $\pm$ 0.05 \\
Star radius [\Rsun] ~$^e$ &  0.79 $\pm$ 0.05  \\
Planet mass $M_{\rm p}$ [\Mjup ]  &  2.75  $\pm$ 0.16 \\
Planet radius $R_{\rm p}$ [\Rjup]  &  0.97 $\pm$ 0.07 \\
Planet density $\rho_{\rm p}$ [$g\;cm^{-3}$] &  3.70 $\pm$ 0.83 \\
Planet surface gravity log\,$g_{\rm p}$ [cgs] &  3.93 $\pm$  0.08 \\
%Stellar rotation period $P_{*, rot}$ [days]  &  27.4 $\pm$ 2.6 \\
Planet rotation period $P_{\rm p, rot}$ [days]  ~$^f$ &  4.25 $\pm$ 0.53 \\
%Age of the star $t$ [Gyr]  ~$^e$ & $ < 3 $ \\
Distance of the star $d$ [pc] & 345 $\pm$ 70 \\
Orbital semi-major axis $a$ [AU] & 0.1055 $\pm$ 0.0021 \\
Orbital distance at periastron $a_{\rm per}$ [AU] & 0.0496 $\pm$ 0.0039 \\
Orbital distance at apoastron $a_{\rm apo}$ [AU] &  0.1614 $\pm$ 0.0047 \\
Equilibrium temperature at the averaged distance
$T_{\rm eq}$ [K] ~$^g$ & 600 $\pm$ 23 \\
Equilibrium temperature at periastron
$T^{\rm per}_{\rm eq}$ [K] ~$^g$ & 935 $\pm$ 54 \\
Equilibrium temperature at apoastron
$T^{\rm apo}_{\rm eq}$ [K] ~$^g$ & 518
$\pm$ 20 \\
%&\\
%\\
\hline       
\vspace{-0.5cm}
\footnotetext[1]{\scriptsize
$T_{\rm occ}= T_{\rm tr}+\frac{P}{\pi} \cdot \left(\frac{\pi}{2}+(1+\csc^2{i})\cdot e~\cos{\omega}\right)$;
}
\footnotetext[2]{\scriptsize 
phase of the  end of transit egress 
in the reference system defined by \citet{GimenezPelayo83}.} 
%in the case of eccentric orbits
% (see \citealt{Gimenez09}, Sect.~2);}
% It allows us to measure
%the true anomaly from conjunction passage
%in the case of eccentric orbits
% (see \citealt{Gimenez09}, Sect.~2); }
 %
%True anomaly measured from conjunction at 
%the transit egress (see Gim09);}
%\footnotetext[2]{\scriptsize $I(\mu)/I(1)=1-u(1-\mu)$, where $I(1)$ is the 
%specific intensity at the center of the disk and $\mu=\cos{\gamma}$,
%$\gamma$ beeing the angle between the surface normal and the line of sight;}
\footnotetext[3]{\scriptsize $a/R_{*}=\frac{1+e \cdot \cos{\nu_{2}}}{1-e^{2}} 
\cdot \frac{1+k}{\sqrt{1-\cos^{2}({\nu_{2}+\omega-\frac{\pi}{2}})
\cdot \sin^{2}{i}}}$,
where $\nu_{2}$ is the true anomaly measured 
from the periastron passage at the  end of transit egress (see \citealt{Gimenez09});}
\footnotetext[4]{\scriptsize $b=\frac{a \cdot \cos{i}}{R_{*}} \cdot \frac{1-e^{2}}{1+e \cdot \sin{\omega}}$;}
\footnotetext[5]{\scriptsize from CESAM stellar evolution models. See Sect.\ref{stellar_planet_param} for details;}
\footnotetext[6]{\scriptsize assuming the planet to be in a pseudo-synchronous rotation;}
%\footnotetext[6]{\scriptsize from gyrochronology \citep{Barnes07};}
%\footnotetext[7]{\scriptsize estimated from $V - M_{V} = 5\log d - 5 + A_{V}$, where $A_{V} = (5.82\pm0.1) \cdot E(J-K)$ 
%\citep{2000asqu.book.....C}; $M_{V}$ and $(J-K)_0$  taken from 
%http://kurucz.harvard.edu for the $Ts_{\rm eff}$ and  $[\rm{Fe/H}]$ of CoRoT-10;}
\footnotetext[7]{\scriptsize black body equilibrium temperature for an isotropic 
planetary emission.}
\end{tabular}
\end{minipage}
\label{starplanet_param_table}  
\end{table*}

Saturated interstellar Na D lines in the \emph{HARPS} spectra indicate a significant absorption along the line of sight.
Converting the 2MASS $J$ and $K$ magnitudes (Table~\ref{startable}) in the Bessel \& 
Brett photometric system \citep{BessellBrett88}, and 
comparing the $(J-K)$ colour with that expected by Kurucz models for the CoRoT-10
spectral type and metallicity, we found a colour excess $E(J-K) \simeq 0.24$. This corresponds to an 
extinction of $A_{V} \simeq 1.39$~mag\footnote{$A_{V}/E(J-K) = 5.82\pm0.1$ 
 \citep{2000asqu.book.....C}}, in agreement
with reddening maps \citep{Schlegeletal98}. Using the Pogson formula\footnote{
$V - M_{V} = 5\log d - 5 + A_{V}$, where $M_{V}$ was determined from 
the bolometric magnitude, given the $T_{\rm eff}$ and $R_{*}$ of CoRoT-10, and the BC taken
from http://kurucz.harvard.edu for the atmospheric parameters of CoRoT-10
}, the stellar distance was estimated to $345 \pm 70$~pc.

%We compared the 2MASS (J-K) colour to the one expected by Kurucz models for a metal
 %rich K1 star and found a colour excess E(J-K) = 0.24. This corresponds to an extinction 
 %of Av = 1.39 mag\footnote{using $Av/E(J-K) = 5.82\pm0.1$ from \citet{2000asqu.book.....C}} 
 %that is consistent with both the extinction maps \citep{} and the crowded line of 
 %sight suggested by strong interstellar absorption Na D lines.

From the transit best-fit we derived a stellar density
of $\rho_{*}=3.32 \pm 0.70~\rm{g\;cm^{-3}}$, i.e. $2.35 \pm 0.50~\rho_{\odot}$. 
CESAM \citep{Morel08} and STAREVOL  \citep{PalaciosPC, Siess06} 
models of stellar evolution  
do not foresee any evolutionary track that matches the 
above-mentioned stellar density,
given the effective temperature and the metallicity of CoRoT-10.
Specifically, they predict an upper-limit of $2.55~\rm {g\;cm^{-3}}$ 
($1.79~\rho_{\odot}$), compatible at 1.1-sigma with 
the stellar density derived
from the transit fitting. 
The mass and radius of the star provided by the CESAM 
evolutionary tracks are respectively equal to $M_{*}=0.89 \pm 0.05~\rm{M_{\odot}}$ 
and $R_{*}=0.79 \pm 0.05~\rm{R_{\odot}}$.
The corresponding surface gravity, 
log\,$g$=$4.59 \pm 0.06~\rm{cm~s^{-2}}$, is in good agreement 
with the spectroscopic value. The stellar age constraints are relatively weak 
but favor values smaller than 3 Gyr.

From the aforementioned stellar radius and mass, we determined the
radius of the planet $R_{\rm p}=0.97 \pm 0.07~\Rjup$ and its mass 
$M_{\rm p}=2.75 \pm 0.16~\Mjup$. The bulk density, 
$\rho_{\rm p}=3.70 \pm 0.83 ~\rm{g\;cm^{-3}}$, is $\sim 2.8$ that of Jupiter.

\section{Discussion and conclusions}
\begin{figure}[b]
\centering
\includegraphics[width=8.5cm]{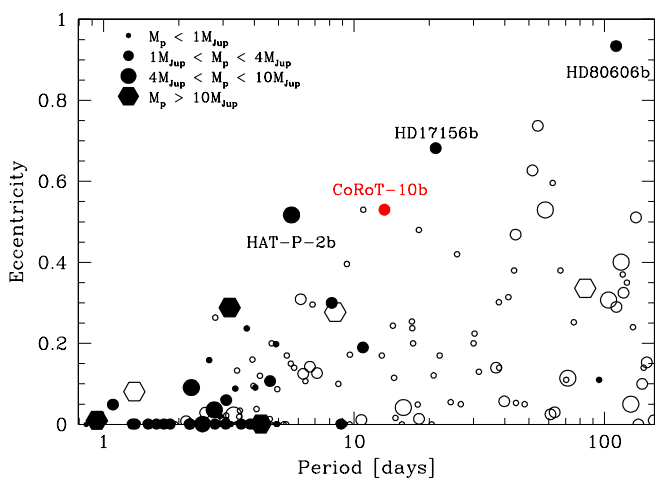}
\caption{Eccentricity - period diagram for the known extrasolar planets 
(black filled symbols are transiting planets). The size of the symbol 
indicates the mass range. CoRoT-10b, with $P=13.24$ days and $e=0.53$, is 
indicated by the red filled circle. Data from http://exoplanet.eu.}
\label{diagram_P_ecc}
\end{figure}

We report the discovery of CoRoT-10b, a transiting planet on a highly eccentric
orbit ($e=0.53 \pm 0.04$) with a mass of $2.75 \pm 0.16 ~\Mjup $ 
and a radius of $0.97 \pm 0.07 ~\Rjup$. It orbits 
a metal rich ($[\rm{Fe/H}]=0.26 \pm 0.07$) K1V star with a visual 
magnitude $V=15.22$ in 13.24 days. Fig.~\ref{diagram_P_ecc} shows the position
of CoRoT-10b in the eccentricity-period diagram 
of the known extrasolar planets and highlights 
its peculiarity as it belongs to the class of the few 
transiting exoplanets with highly eccentric orbits ($e \gtrsim 0.5$) 
among which HAT-P-2b, HD~17156b and HD~80606b.  
%CoRoT-10b has both the smallest mass and radius of them.

\begin{figure}[t]

\centering
\includegraphics[width=9.0cm]{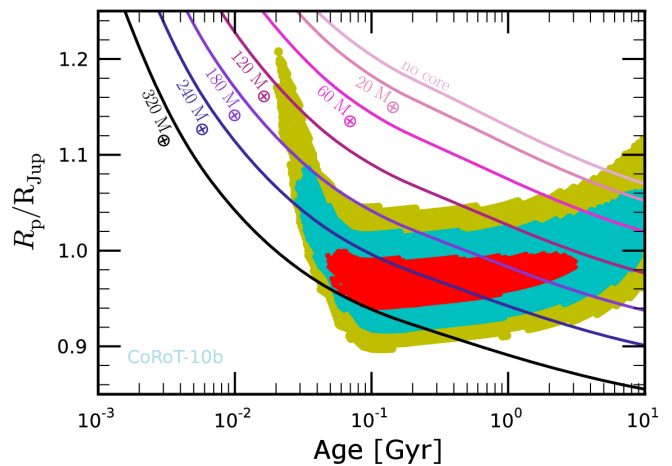}
\caption{Evolution of the radius of CoRoT-10b (in Jupiter units) as a function of 
age, compared to
constraints inferred from CoRoT photometry, spectroscopy, radial velocimetry and
CESAM stellar evolution models. Red, blue and green areas correspond to the
planetary radii and ages that result from stellar evolution models matching the
inferred $\rho_{*}-T_{\rm eff}$ uncertainty ellipse within 1$\sigma$,
2$\sigma$ and 3$\sigma$, respectively. Planetary evolution models for a planet
with a solar-composition envelope over a central dense core of pure
rocks of variable mass are shown as solid lines. These models assume a total
mass of 2.75\,$\Mjup$ and an equilibrium temperature of 600\,K
(corresponding to a zero Bond albedo). They depend weakly on the assumed opacities, and
uncertainties due to the atmospheric temperature and planetary mass are
negligible.\label{mathieu}}
\end{figure}

% structure of the planet {\bf TO BE DONE WITH THE HELP OF TRISTAN} \\
To investigate the internal structure of CoRoT-10b, we computed
planetary evolution models with CEPAM \citep{Guillot95} under the standard
hypothesis that the planet is made up of a central
rocky core of variable mass and of an overlying envelope of solar
composition \citep[e.g.,][]{Guillot08}. The results in terms of planetary size as a function of system
age are shown in Fig.~\ref{mathieu}. 
The coloured regions (red, blue, yellow-green) indicate the constraints
derived from the stellar evolution models at 1, 2
and 3 $\sigma$ level, respectively. Assuming a zero Bond albedo, we derived the equilibrium 
temperature of the planet $T_{\rm eq}=600$~K. For this temperature,
models of planet internal structure with a core mass of 0, 20, 60, 
120, 180, 240 and 320 M$_\oplus$ were computed 
(Fig.~\ref{mathieu}). Note that
a 25\% change in the equilibrium temperature yields 
a difference in the resulting planetary radius of less than 1\%.
Therefore, a Bond albedo considerably different
from zero (up to $A_{\rm B}=0.7$) does not change significantly our results.

CoRoT-10b is a high density planet with a mass and density similar to those of
HD~17156b \citep{Barbierietal09} but with a higher content of heavy elements. For 
an age of the star and hence of the planetary system between 1 and 3 Gyr, 
CoRoT-10b should contain 
between 120 and 240 M$_\oplus$ of rocks in
its interior (i.e. between 14 and 28\% of the total mass), at 1-sigma level.
Mixing heavy elements in the envelope rather than assuming that they are 
all contained in the core may yield a reduction of these numbers by $\sim 30\%$  
\citep{Baraffeetal08}. This number is uncertain however because it does not 
account for the increase in opacity in the outer radiative zone that would have 
the opposite effect \citep{Guillot05}. In any case, CoRoT-10b is found to be 
extremely enriched in heavy elements, suggesting that its formation probably 
required giant collisions (see \citealt{Ikomaetal06}). It also strengthens the 
observed correlation between star metallicity and heavy elements in the planet \citep{Guillot08}.

%INSOLATION, secondary eclipse, rotation period of the planet, model of Iro & Deming, Sudarsky models?
Moving along its eccentric orbit from the periastron to the apoastron,
CoRoT-10b experiences a 10.6-fold variation in insolation. Moreover, since 
the tidal interaction with the parent star is strongest 
around periastron, the planet is expected to be in pseudo-synchronous 
rotation. Eq.~42 from \citet{Hut81} gives a rotation period of the planet 
$P_{\rm p, rot}= 4.25 \pm 0.53$ days. Time-dependent 
radiative models of planetary
atmospheres could be used to study heating variations 
due to the changing star-planet distance at different
pressure levels of the planet atmosphere and to predict temperature 
inversions caused by strong heating around periastron, as done
by \citet{IroDeming10} for HD~17156b and HD~80606b.

%Very recently, \citet{IroDeming10} presented
%a time-dependent radiative model for the atmosphere of eccentric
%planets, taking into account their pseudo-synchronization. As for
%HD~17156b and HD~80606b, such a model could be applied to 
%CoRoT-10b in order to a) study heating variations 
%due to the changing star-planet distance at different
%pressure levels of the planetary atmosphere; and b) predict temperature 
%inversions caused by strong heating 
%around periastron \citet{IroDeming10}. 
%and c) predict longitudinal shifts between the substellar 
%point and the hottest zone of the atmosphere because of both the finite 
%radiative response times and the planet rotation (Iro \& Deming, 2010). 

% Tidal dissipation, planet-planet scattering, Kozai mechanism, leggere proposal UVES
%Using Eq.~1 in \citet{Matsumuraetal08}, we can estimate the circularisation time $\tau_{circ}$ 
%of the planetary orbit  that strongly depends on the adopted 
%tidal quality factor for the planet 
%$Q'_{p}$ and its parent star $Q'_{s}$, which are not well known. For Jupiter, 
%$5 \cdot 10^{4}<Q'_{p}<2 \cdot 10^{6}$ \citep{YoderPeale81}
 %and for most of the giant extrasolar planets $10^{5}<Q'_{p}<10^{9}$ 
 %(see Fig.~2 from \citealt{Matsumuraetal08}). Considering
% $Q'_{p}=10^{5}$ and $Q'_{s}=10^{6}$,
% we find  $\tau_{circ} \sim 6.9$ Gyr. Since $\tau_{circ}$ is longer than the stellar age,
% the eccentricity of CoRoT-10b need not be excited and 
% maintained by the resonant
% interaction with another planet. 

Using Eq.~6 in \citet{Matsumuraetal08}, we can estimate the circularisation time $\tau_{\rm circ}$ 
of the planetary orbit, neglecting the stellar damping.  The circularisation time
 strongly depends on the adopted 
tidal quality factor for the planet 
$Q'_{\rm p}$ which is not well known. For Jupiter, 
$5 \cdot 10^{4}<Q'_{\rm p}<2 \cdot 10^{6}$ \citep{YoderPeale81, Laineyetal09}
 and for most of the giant extrasolar planets $10^{5}<Q'_{\rm p}<10^{9}$ 
 (see Fig.~2 from \citealt{Matsumuraetal08}). Considering
 $Q'_{\rm p}=10^{5}$,
 we find  $\tau_{\rm circ} \sim 7.4$~Gyr. Owing to the long circularisation time,
 the eccentricity of CoRoT-10b need not be excited and 
 maintained by the resonant
 interaction with another planet. 
 Nevertheless, the  
 eccentric orbits of extrasolar planets can be explained 
 by gravitational planet-planet scattering (e.g., \citealt{Chatterjeeal08}
 and references therein). 
 If another massive planet survived the violent encounters between planets 
 and is currently orbiting around the parent star,
 it could be detected by a long-term radial velocity follow-up of the parent star,
% with the \emph{HARPS} spectrograph.
 e.g., showing 
 %More radial velocity observations might also reveal 
 a long-term drift
 induced by the distant companion. 
 If the star has a distant companion
 of stellar nature, 
% This would suggest 
 the high eccentricity of CoRoT-10b could be produced by Kozai
 oscillations rather than planet-planet scattering (e.g., \citealt{TakedaRasio05}).
Distinguishing between the two scenarios would make us
understand the dynamical evolution of the eccentric
giant planet CoRoT-10b.

%We searched for the presence of the secondary transit in the CoRoT data
%at the expected phase of 0.238 using the same procedures as in 
%Fridlund et al. 2010 but we were not able to detect it 
%for two reasons: first, the considerable star-planet distance 
% $r_{t}=0.0557 AU$ at the occultation of the planet after the
%periastron passage; and secondly, the low precision 

% Radial velocity FU 

% Additional companion
%The transit of CoRoT-10b is also highly grazing: the impact parameter
%$b=0.89 \pm 0.01$ is equal to that of HAT-P-14b (ref. ?).

% Rossiter MC UVES

\begin{acknowledgements}
The authors wish to thank the staff at ESO La Silla Observatory for their support and for their contribution to the success of the HARPS project and operation. The team at IAC acknowledges support by grant ESP2007-65480-C02-02 of the Spanish Ministerio de Ciencia e Innovaci\'on. The CoRoT/Exoplanet catalogue (Exodat) was made possible by observations collected for years at the Isaac Newton Telescope (INT), operated on the island of La Palma by the Isaac Newton group in the Spanish Observatorio del Roque de Los Muchachos of the Instituto de Astrophysica de Canarias. The German CoRoT team (TLS and University of Cologne) acknowledges DLR grants 50OW0204, 50OW0603, and 50QP0701. The French team wish to thank the ÓProgramme National de Plan\'etologieÓ (PNP) of CNRS/INSU and the French National Research Agency (ANR-08- JCJC-0102-01) for their continuous support to our planet search. The Swiss team acknowledges the ESA PRODEX program and the Swiss National Science Foundation for their continuous support on CoRoT ground follow-up. 
A. S. Bonomo acknowledges CNRS/CNES grant 07/0879-Corot.
S. Aigrain acknowledges STFC grant ST/G002266.
M. Gillon acknowledges support from the Belgian Science Policy Office in the form of a Return Grant. 
 
\end{acknowledgements}

~\\
~\\


\begin{thebibliography}{}
\bibliographystyle{aa}

{\small

\bibitem[\protect\citeauthoryear{Aigrain et al.} {2008}]{Aigrainetal08}
Aigrain, S., Collier Cameron, A., Ollivier, M., et al. 2008, \aap, 488, L43

\bibitem[\protect\citeauthoryear{Alonso et al.} {2008a}]{Alonsoetal08a}
Alonso, R., Auvergne, M., Baglin, A., et al. 2008a, \aap, 482, L21

\bibitem[\protect\citeauthoryear{Alonso et al.} {2008b}]{Alonsoetal08b}
Alonso, R., Barbieri, M., Rabus, M., et al. 2008b, \aap, 487, L5

\bibitem[\protect\citeauthoryear{Auvergne et al.} {2009}]{Auvergneetal09}
Auvergne, M., Bodin, P., Boisnard, L., et al. 2009, \aap, 506, 411


\bibitem[\protect\citeauthoryear{Baglin} {2003}]{Baglin03} 
Baglin, A. 2003, Adv. Sp. Res., 31, 345

%\bibitem[{Baglin}{2003}] {Baglin03} 
%Baglin, A. 2003, Adv. Sp. Res., 31, 345

\bibitem[\protect\citeauthoryear{Bakos et al.} {2007}]{Bakosetal07}
Bakos, G., \'A., Kov\'acs, G., Torres, G., et  al. 2007, \apj, 670, 826

\bibitem[\protect\citeauthoryear{Bakos et al.} {2010}]{Bakosetal10}
Bakos, G., \'A., Torres, G., P\'al, A., et  al. 2010, \apj, 710, 1724

%\bibitem[\protect\citeauthoryear{Baliunas et al.} {1996}]{Baliunas1996} 
%Baliunas, S., Sokoloff, D., \& Soon, W.\ 1996, \apjl, 457, L99 


\bibitem[\protect\citeauthoryear{Baraffe et al.} {2008}]{Baraffeetal08} 
Baraffe, I., Chabrier, G. \& Barman, T. 2008, \aap, 482, 315

\bibitem[\protect\citeauthoryear{Baranne} {1996}]{baranne96} 
Baranne, A., Queloz, D., Mayor, M., et al. 1994, \aaps, 119, 373

%\bibitem[\protect\citeauthoryear{Barnes} {2007}]{Barnes07} 
%Barnes, S. A. 2009, \apj, 669, 1189

\bibitem[\protect\citeauthoryear{Barbieri et al.} {2007}]{Barbierietal07}
Barbieri, M., Alonso, R., Laughlin, G. et al. 2007, \aap, 476, L13

\bibitem[\protect\citeauthoryear{Barbieri et al.} {2009}]{Barbierietal09}
Barbieri, M., Alonso, R., Desidera, S. et al. 2009, \aap, 503, 601

\bibitem[\protect\citeauthoryear{Barge et al.} {2008}]{Bargeetal08}
Barge, P., Baglin, A., Auvergne, M., et al. 2008, \aap, 482, L17

\bibitem[\protect\citeauthoryear{Bean et al.} {2008}]{Beanetal08}
Bean, J. L., Benedict, G. F., Charbonneau, D., et al. 2008, \aap, 486, 1039

%\bibitem[\protect\citeauthoryear{Bouchy et al.} {2008}]{bouchy08} 
%Bouchy, F., Moutou, C., Queloz, D., et al., 2008, in 
%Transiting Planets, Proceedings of the IAU Symposium, Vol. 253, 129

%\bibitem[\protect\citeauthoryear{Boisse et al.} {2009}]{Boisse2009} 
%Boisse, I., et al.\ 2009, \aap, 495, 959

\bibitem[\protect\citeauthoryear{Bessell \& Brett} {1988}]{BessellBrett88}
Bessel, M. S. \& Brett, J. M. 1988, \pasp, 100, 1134

%\bibitem[\protect\citeauthoryear{Bodenheimer et al.} {2003}]{Bodenheimer03}
%Bodenheimer, P., Laughlin, G. \& Lin, D. N. C. 2003, \apj, 592, 555

\bibitem[\protect\citeauthoryear{Bord\'e et al.} {2010}]{Bordeetal10}
Bord\'e, P., Bouchy, F., Deleuil, M., et al. 2010, \aap, accepted

\bibitem[\protect\citeauthoryear{Bouchy et al.} {2008}]{bouchy08} 
Bouchy, F., Moutou, C., Queloz, D., et al. 2008, in 
Transiting Planets, Proceedings of the IAU Symposium, Vol.~253, p.~129

%\bibitem[\protect\citeauthoryear{Brown et al.} {2001}]{Brownetal01}
%Brown, T. M., Charbonneau, D., Gilliland, R. L., et al. 2001, \apj, 552, 699

\bibitem[\protect\citeauthoryear{Bruntt} {2009}]{Bruntt2009} 
Bruntt, H.\ 2009, \aap, 506, 235 

\bibitem[\protect\citeauthoryear{Cabrera et al.} {2009}]{Cabreraetal09}
Cabrera, J., Fridlund, M., Ollivier, M., et al. 2009, \aap, 506, 501

%\bibitem[\protect\citeauthoryear{Cameron et al.} {2009}] {Cameronetal09}
%Cameron, A. C., Davidson, V., A., Hebb, L., et al. 2010, \mnras, in press

\bibitem[\protect\citeauthoryear{Chatterjee et al.} {2008}] {Chatterjeeal08}
Chatterjee, S., Ford, E. B., Matsumura, S. \& Rasio, F. A. et al. 2008, \apj, 685, 580

\bibitem[\protect\citeauthoryear{Cox} {2000}]{2000asqu.book.....C} 
Cox, A.~N.\ 2000, Allen's Astrophysical Quantities

\bibitem[\protect\citeauthoryear{Deeg et al.} {2009}]{Deegetal09}
Deeg, H., Gillon, M., Shporer, A., et al. 2009, \aap, 506, 343

\bibitem[\protect\citeauthoryear{Deeg et al.} {2010}]{Deegetal10}
Deeg, H., Moutou, C., Erikson, A., et al. 2010, Nature, 464, 384

\bibitem[\protect\citeauthoryear{Deleuil et al.} {2008}]{Deleuiletal08}
Deleuil, M., Deeg, H. J., Alonso, R., et al. 2008, \aap, 491, 889

\bibitem[\protect\citeauthoryear{Deleuil et al.} {2009}]{Deleuiletal09}
Deleuil, M., Meunier, J. C., Moutou, C., et al. 2009, \aj, 138, 649

\bibitem[\protect\citeauthoryear{Fridlund et al.} {2010}]{Fridlundetal10}
Fridlund, M., H\'ebrard, G., Alonso, R. et al. 2010, \aap, in press, arXiv:1001.1426

%\bibitem[\protect\citeauthoryear{Gazzano et al.} {2010}]{Gazzano2010} 
%Gazzano, J.-C.\ 2010, in preparation.

\bibitem[\protect\citeauthoryear{Gillon et al.} {2007a}]{Gillonetal07a}
Gillon, M., Pont, F., Demory B.-O., et al. 2007a, \aap, 472, L13

\bibitem[\protect\citeauthoryear{Gillon et al.} {2007b}]{Gillonetal07b}
Gillon, M., Demory B.-O., Barman T., et al. 2007b, \aap, 471, L51

\bibitem[\protect\citeauthoryear{Gim\'enez} {2006}]{Gimenez06}
Gim\'enez, A. 2006, \aap, 450, 1231

\bibitem[\protect\citeauthoryear{Gim\'enez} {2009}]{Gimenez09}
Gim\'enez, A. 2009, in The Eighth Pacific Rim Conference on Stellar 
Astrophysics: A Tribute to Kam Ching Leung,
Eds. B. Soonthornthum, S. Komonjinda, K.S. Cheng,
and K.C. Leung San Francisco: Astronomical Society of the Pacific, 
Vol.~450, p.~291

\bibitem[\protect\citeauthoryear{Gim\'enez \& Garcia-Pelayo} {1983}]{GimenezPelayo83}
Gim\'enez, A. \& Garcia-Pelayo, J. M. 1983, \apss, 92, 203

\bibitem[\protect\citeauthoryear{Guillot \& Morel} {1995}]{Guillot95}
Guillot, T. \& Morel, P. 1995, \aaps, 109, 109

\bibitem[\protect\citeauthoryear{Guillot} {2005}]{Guillot05}
Guillot, T. 2005, Annual Review of Earth and Planetary Sciences, Vol.~33, p.~493

\bibitem[\protect\citeauthoryear{Guillot} {2008}]{Guillot08}
Guillot, T. 2008, Physica Scripta Volume T, 130, 014023

%\bibitem[\protect\citeauthoryear{Hall et al.} {2007}]{Hall2007} 
%Hall, J.~C., Lockwood, G.~W., \& Skiff, B.~A.\ 2007, \aj, 133, 862 

\bibitem[\protect\citeauthoryear{H\'ebrard et al.} {2010}]{Hebrardetal10}
H\'ebrard, G., Desert, M.-J., D\'iaz, R. F. et al. 2010, \aap, in press, arXiv:1004.0790

\bibitem[\protect\citeauthoryear{Hut} {1981}]{Hut81}
Hut, P. 1981, \aap, 99, 126

\bibitem[\protect\citeauthoryear{Ibgui et al.} {2010}]{Ibgui10}
Ibgui, L., Burrows, A. \& Spiegel, D. S. 2010, \apj, 713, 751 

\bibitem[\protect\citeauthoryear{Ikoma et al.} {2006}]{Ikomaetal06}
Ikoma, M., Guillot, T., Genda, H. et al. 2006, \apj, 650, 1150

\bibitem[\protect\citeauthoryear{Iro \& Deming} {2010}]{IroDeming10}
Iro, N. \& Deming, L. D. 2010, \apj, 712, 218

\bibitem[\protect\citeauthoryear{Johns-Krull et al.} {2008}]{JohnsKrulletal08}
Johns-Krull, C. M., McCullough, P. R., Burke, C. J., et al. 2008, \apj, 677, 657

\bibitem[\protect\citeauthoryear{Kov\'acs et al.} {2010}]{Kovacsetal10}
Kov\'acs, G., Bakos, G. \'A., Hartman, J. D., et al. 2010, \apj, submitted, arXiv:1005.5300

\bibitem[\protect\citeauthoryear{Kozai} {1962}]{Kozai62}
Kozai, Y. 1962, \aj, 67, 591

\bibitem[\protect\citeauthoryear{Lainey et al.} {2009}]{Laineyetal09}
Lainey, V., Arlot, J.-E., Karatekin, \"O. \& Van Hoolst, T. 2009, Nature, 459, 957

\bibitem[\protect\citeauthoryear{Langton \& Laughlin} {2007}]{Langton07}
Langton, J. \& Laughlin, G. 2007, \apj, 657, L113

\bibitem[\protect\citeauthoryear{Lanza et al.} {2009}]{Lanzaetal09}
Lanza, A. F., Pagano, I., Leto, G., et al. 2009, \aap, 493, 193

\bibitem[\protect\citeauthoryear{Lanza et al.} {2010}]{Lanzaetal10}
Lanza, A. F., Bonomo, A. S., Moutou, C., et al. 2010, \aap, in press, arXiv:1005.3602

\bibitem[\protect\citeauthoryear{L\'eger et al.} {2009}]{Legeretal09}
L\'eger, A., Rouan, D., Schneider, J., et al. 2009, \aap, 506, 287

%\bibitem[\protect\citeauthoryear{Mamajek \& Hillenbrand} {2008}]{Mamajek2008} 
%Mamajek, E.~E., \& Hillenbrand, L.~A.\ 2008, \apj, 687, 1264

\bibitem[\protect\citeauthoryear{Marzari \& Weidenschilling} {2002}]{Marzari02} 
Marzari, F. \& Weidenschilling, S. J. 2002, Icarus, 156, 570

\bibitem[\protect\citeauthoryear{Matsumura et al.} {2008}]{Matsumuraetal08} 
Matsumura, S., Takeda, G. \& Rasio, F. A. et al. 2008, \apj, 686, L29

\bibitem[\protect\citeauthoryear{Mayor et al.} {2003}]{mayor03} 
Mayor, M., Pepe, F., Queloz, D., et al. 2003, The Messenger, 114, 20

\bibitem[\protect\citeauthoryear{Morel} {2008}]{Morel08}
Morel, P. \& Lebreton, Y. 2008, \apss, 316, 61

\bibitem[\protect\citeauthoryear{Moutou et al.} {2008}]{Moutouetal08}
Moutou, C., Bruntt, H., Guillot, T., et al. 2008, \aap, 488, L47

\bibitem[\protect\citeauthoryear{Moutou et al.} {2009}]{Moutouetal09}
Moutou, C., H\'ebrard, G., Bouchy, F., et al. 2009, \aap, 498, L5

%\bibitem[\protect\citeauthoryear{Noyes et al.} {1984}]{Noyes1984} 
%Noyes, R.~W., Hartmann, L.~W., Baliunas, S.~L., Duncan, D.~K., \& Vaughan, A.~H.\ 1984, \apj, 279, 763

\bibitem[\protect\citeauthoryear{P\'al et al.} {2010}]{Paletal10}
P\'al, A., Bakos, G. \'A., Torres, G., et  al. 2010, \mnras, 401, 2665

\bibitem[\protect\citeauthoryear{Palacios,} {private communication}]{PalaciosPC}
Palacios, A., Calculation of dedicated evolutionary tracks using STAREVOL, private communication.

\bibitem[\protect\citeauthoryear{Pepe et al.} {2002a}]{pepe02a} 
Pepe, F., Mayor, M., Galland, F., et al. 2002a, \aap, 388, 632

\bibitem[\protect\citeauthoryear{Pepe et al.} {2002b}]{pepe02b} 
Pepe, F., Mayor, M., Ruppretch, G., et al. 2002b, Messenger, 110, 9

\bibitem[\protect\citeauthoryear{Press et al.} {1992}]{Pressetal92} 
Press, W. H., Teukolsky, S. A., Vetterling, W. T., Flannery, B. P. 1992,
Numerical recipes in FORTRAN, The art of scientific computing (Cambridge:
University Press), 2nd edn

\bibitem[\protect\citeauthoryear{Queloz et al.} {2008}]{Quelozetal09}
Queloz, D., Bouchy, F., Moutou, C., et al. 2009, \aap, 506, 303

 \bibitem[\protect\citeauthoryear{Queloz et al.} {2010}]{Quelozetal10}
Queloz, D., Anderson, D., Cameron, A. C. et al. 2010, \aap, submitted

\bibitem[\protect\citeauthoryear{Rauer et al.} {2009}]{Raueretal09}
Rauer, H., Queloz, D., Csizmadia, S., et al. 2009, \aap, 506, 281

\bibitem[\protect\citeauthoryear{Recio-Blanco et al.} {2006}]{Recio-Blanco2006} 
Recio-Blanco, A., Bijaoui, A., \& de Laverny, P.\ 2006, \mnras, 370, 141 

\bibitem[\protect\citeauthoryear{Scargle} {1982}]{Scargle82}
Scargle, J. D. 1982, \apj, 263, 835

\bibitem[\protect\citeauthoryear{Schlegel et al.} {1998}]{Schlegeletal98}
Schlegel, D. J., Finkbeiner, D. P. \& Davis, M. 1998, \apj, 500, 525

\bibitem[\protect\citeauthoryear{Siess} {2006}]{Siess06}
Siess, L. 2006, \aap, 448, 717

\bibitem[\protect\citeauthoryear{Sing} {2010}]{Sing10}
Sing, D. K. 2010, \aap, 510, A21

\bibitem[\protect\citeauthoryear{Surace et al.} {2008}]{Suraceetal08}
Surace, C., Alonso, R., Barge, P., et al. 2008, in Presented at the Society of Photo-
Optical Instrumentation Engineers (SPIE) Conference, Vol. 7019, Society of
Photo-Optical Instrumentation Engineers (SPIE) Conference Series

\bibitem[\protect\citeauthoryear{Takeda \& Rasio} {2005}]{TakedaRasio05}
Takeda, G. \& Rasio, F. A. 2005, \apj, 627, 1001

\bibitem[\protect\citeauthoryear{Yoder \& Peale} {1981}]{YoderPeale81}
Yoder, C. F. \& Peale, S. J. 1981, Icarus, 47, 1

\bibitem[\protect\citeauthoryear{Winn et al.} {2008}]{Winnetal08}
Winn, J. N., Holman, M. J. \& Torres, G., et al. 2008, \apj, 683, 1076

\bibitem[\protect\citeauthoryear{Winn et al.} {2009}]{Winnetal09}
Winn, J. N., Howard, A. W., Johnson, J. A., et al. 2009, \apj, 703, 2091

%\bibitem[\protect\citeauthoryear{Zechmeister \& K\"{u}rster} {2009}]{ZechmeisterKurster09}
%Zechmeister, M. \& K\"{u}rster, M., 2009, \aap, 496, 577

}

\end{thebibliography}
\end{document}